\documentstyle[epsfig]{mn}

\newif\ifAMStwofonts

\def\gtorder{\mathrel{\raise.3ex\hbox{$>$}\mkern-14mu
             \lower0.6ex\hbox{$\sim$}}}
\def\ltorder{\mathrel{\raise.3ex\hbox{$<$}\mkern-14mu
             \lower0.6ex\hbox{$\sim$}}}

\ifoldfss
  \ifCUPmtlplainloaded \else
    \NewTextAlphabet{textbfit} {cmbxti10} {}
    \NewTextAlphabet{textbfss} {cmssbx10} {}
    \NewMathAlphabet{mathbfit} {cmbxti10} {} 
    \NewMathAlphabet{mathbfss} {cmssbx10} {} 
  \fi
  \ifAMStwofonts
    \ifCUPmtlplainloaded \else
      \NewSymbolFont{upmath} {eurm10}
      \NewSymbolFont{AMSa} {msam10}
      \NewMathSymbol{\upi}     {0}{upmath}{19}
      \NewMathSymbol{\umu}     {0}{upmath}{16}
      \NewMathSymbol{\upartial}{0}{upmath}{40}
      \NewMathSymbol{\leqslant}{3}{AMSa}{36}
      \NewMathSymbol{\geqslant}{3}{AMSa}{3E}

       \let\le=\leqslant
       \let\ge=\geqslant
    \fi
  \fi
\fi 

\ifnfssone
  \newmathalphabet{\mathit}
  \addtoversion{normal}{\mathit}{cmr}{m}{it}
  \addtoversion{bold}{\mathit}{cmr}{bx}{it}
  \newmathalphabet{\mathbfit} 
  \addtoversion{normal}{\mathbfit}{cmr}{bx}{it}
  \addtoversion{bold}{\mathbfit}{cmr}{bx}{it}
  \newmathalphabet{\mathbfss} 
  \addtoversion{normal}{\mathbfss}{cmss}{bx}{n}
  \addtoversion{bold}{\mathbfss}{cmss}{bx}{n}
  \ifAMStwofonts
    \ifCUPmtlplainloaded \else
      \UseAMStwoboldmath
      \makeatletter
      \new@mathgroup\upmath@group
      \define@mathgroup\mv@normal\upmath@group{eur}{m}{n}
      \define@mathgroup\mv@bold\upmath@group{eur}{b}{n}
      \edef\UPM{\hexnumber\upmath@group}
      \new@mathgroup\amsa@group
      \define@mathgroup\mv@normal\amsa@group{msa}{m}{n}
      \define@mathgroup\mv@bold\amsa@group{msa}{m}{n}
      \edef\AMSa{\hexnumber\amsa@group}
      \makeatother
      \mathchardef\upi="0\UPM19
      \mathchardef\umu="0\UPM16
      \mathchardef\upartial="0\UPM40
      \mathchardef\leqslant="3\AMSa36
      \mathchardef\geqslant="3\AMSa3E

       \let\le=\leqslant
       \let\ge=\geqslant
    \fi
  \fi
\fi 

\ifnfsstwo
  \DeclareMathAlphabet{\mathbfit}{OT1}{cmr}{bx}{it}
  \SetMathAlphabet\mathbfit{bold}{OT1}{cmr}{bx}{it}
  \DeclareMathAlphabet{\mathbfss}{OT1}{cmss}{bx}{n}
  \SetMathAlphabet\mathbfss{bold}{OT1}{cmss}{bx}{n}
  \ifAMStwofonts
    \ifCUPmtlplainloaded \else
      \DeclareSymbolFont{UPM}{U}{eur}{m}{n}
      \SetSymbolFont{UPM}{bold}{U}{eur}{b}{n}
      \DeclareSymbolFont{AMSa}{U}{msa}{m}{n}
      \DeclareMathSymbol{\upi}{0}{UPM}{"19}
      \DeclareMathSymbol{\umu}{0}{UPM}{"16}
      \DeclareMathSymbol{\upartial}{0}{UPM}{"40}
      \DeclareMathSymbol{\leqslant}{3}{AMSa}{"36}
      \DeclareMathSymbol{\geqslant}{3}{AMSa}{"3E}

       \let\le=\leqslant
       \let\ge=\geqslant
    \fi
  \fi
\fi 

\ifCUPmtlplainloaded \else
  \ifAMStwofonts \else 
    \def\upi{\pi}
    \def\umu{\mu}
    \def\upartial{\partial}
  \fi
\fi

\title[The Type-Ia Supernova Rate in $z \le 1$ Galaxy Clusters: 
Progenitors and the Source of Iron] {The Type-Ia Supernova Rate in $z \le 1$ Galaxy Clusters: 
Implications for Progenitors and the Source of Cluster Iron}

\date{Accepted - .
      Received - ;}

\author[D. Maoz and A. Gal-Yam]{Dan Maoz$^{1}$ and Avishay Gal-Yam$^{1,2}$\\
$^{1}$ School of Physics \& Astronomy and Wise Observatory, 
Tel Aviv University, Tel Aviv 69978, Israel; 
dani@wise.tau.ac.il; avishay@wise.tau.ac.il\\
$^{2}$ Colton Fellow. \\}

\begin{document}

\maketitle


\begin{abstract}

The iron mass in galaxy clusters is about 6 times larger than 
could have been produced by core-collapse supernovae (SNe), 
assuming the stars in the cluster formed with a standard initial 
mass function (IMF). SNe~Ia have been proposed as the
alternative dominant iron source.  Different SN~Ia progenitor models
predict different ``delay functions'', between the formation of a 
stellar population and the explosion of some of its members as SNe~Ia.
We use our previous measurements of the cluster SN~Ia rate at high 
redshift to constrain SN~Ia progenitor models and the
star-formation epoch in clusters.
The low observed rate of cluster SNe~Ia at $z\sim0 - 1$ means that, 
if SNe~Ia produced the observed amount of iron, they must have exploded 
at even higher $z$. This puts a $>95\%$ upper limit on the mean SN~Ia 
delay time of $\tau<2$~Gyr ($<5$~Gyr) if the stars in clusters  
formed at $z_f<2$ ($z_f<3$), assuming $H_{o}=70$ km s$^{-1}$ Mpc$^{-1}$. 
In a companion paper, we show that, for some current versions of cosmic
(field) star formation history (SFH), 
observations of field SNe~Ia place a {\it lower} bound on the delay time,
$\tau>3$~Gyr. If these SFHs are confirmed, the entire range of $\tau$ will
be ruled out. Cluster enrichment by core-collapse SNe
from a top-heavy IMF will then remain the only viable option.

\end{abstract}

\begin{keywords}
supernovae: general
\end{keywords}

\section{Introduction}

Galaxy clusters, by virtue of their deep gravitational potentials, are
``closed boxes'', from which little matter can escape. They thus 
constitute ideal sites to study the time-integrated enrichment of the
intergalactic medium (see e.g., Buote 2002, for a recent review). 
Observations of galaxy clusters have revealed a number of intriguing puzzles.
One such puzzle follows from X-ray spectroscopy, and shows that 
the intracluster medium (ICM) gas consistently 
has a surprisingly high iron abundance, with a ``canonical''
value of about 0.3 the Solar abundance (e.g., Mushotzky \& Loewenstein 1997;
Fukazawa et al. 1998; Finoguenov et al. 2000; White 2000).
Combined with the fact that most of the baryonic mass of the cluster is in
the ICM, this translates into a large mass of iron. 
The tight correlation between total iron mass and stellar light from
the early-type galaxy population (and the lack of a correlation with
late-type galaxies) suggest that the stellar population, whose remnants
and survivors  now populate the early-type galaxies, produced the ICM iron
(Renzini et al. 1993; Renzini 1997).
The near constancy of the ICM iron abundance with cluster mass (Renzini 1997;
Lin, Mohr, \& Stanford 2003) and with redshift out to $z\sim 1$
(Tozzi et al. 2003)
further argues against a significant role in ICM iron enrichment for recent 
infall and disruption of metal-rich dwarf galaxies.
However, the iron mass is at least several times 
larger than that expected from core-collapse supernovae (SNe),
based on the present-day stellar masses, and assuming a standard stellar 
initial mass function (IMF; e.g.,  Renzini et al. 
1993;  Loewenstein 2000). The problem
is aggravated if one considers the large mass of iron that exists in the
cluster galaxies themselves.

A possibly related problem is the 
energy budget of the ICM gas and the ``entropy floor'' observed in clusters,
which suggest a non-gravitational energy source to the ICM, again 
several times larger than the expected energy input from core-collapse SNe
(e.g., Lloyd-Davies, Ponman, \& Cannon 2000; Tozzi \& Norman 2001; 
Brighenti \& Mathews 2001; Pipino et al. 2002). 

Proposed solutions to these problems have included an IMF skewed toward 
high-mass stars (so that a large number of iron-enriching core-collapse 
SNe are produced per present-day unit stellar luminosity), or a dominant 
role for SNe~Ia in the ICM  Fe enrichment. For example, Brighenti \&
Mathews (1998) calculate that the cluster SN~Ia rate must be 
$> 4.8 h^{2}$~SNu  today and
$> 9.6 h^{2}$~SNu at $z=1$ to explain production
of most of the iron in the ICM with SNe~Ia.
[1~SNu$=1~{\rm SN~century}^{-1}(10^{10}
L_{B\odot})^{-1}$.]
This contrasts with the local elliptical-galaxy SN rate of
$(0.28\pm0.12)h^2$~SNu (Cappellaro et al. 1999) and argues against
the type-Ia enrichment scenario. On the other hand, Renzini (1997)
has pointed out that the approximately Solar abundance ratios measured in 
cluster ellipticals argue for a mix of SN-types, and hence also an IMF,
that are not too different from the mix
that exists in the Milky Way. Attempts to derive the SN mix in
clusters by means of direct X-ray measurements of element abundance
ratios in the ICM are still ambiguous, with some results favoring
a dominance of Ia's (e.g., Buote 2002; Tamura et al. 2002) 
and others a dominance
of core-collapse SNe (e.g., Lima Neto et al. 2003; Finoguenov, Burkert,
\& Boehringer 2003).

In Gal-Yam, Maoz, \& Sharon (2002), we used multiple deep
{\it Hubble Space Telescope (HST)} archival images of galaxy clusters
to discover distant field and cluster SNe. The sample was composed
of rich clusters, with X-ray temperatures in the range $4-12$~keV,
and a median of 9~keV. 
The candidate type-Ia SNe in the clusters then led to an estimate of
the SN~Ia rate in the central $250 h^{-1}$~kpc of
medium-redshift $(0.18 \le z \le 0.37)$ and high-redshift
$(0.83\le z \le 1.27)$ cluster sub-samples.
The measured rates are low. To within errors, they are not different from
SN~Ia measurements in field environments,
both locally (Cappellaro et al. 1999) and at high redshift (Pain et al. 2002;
Tonry et al. 2003). It was argued that our 95 per cent upper limits
on the cluster SN~Ia rates rule out the particular model by Brighenti \&
Mathews (1998) for SNe~Ia as the primary source of iron in the ICM.

The issue of SN~Ia rate vs cosmic time 
is closely tied to the presently unsolved question regarding the
progenitor populations of SNe~Ia. Different models 
predict different delay times between the formation of a stellar
population and the explosion of some of its members as SNe~Ia
(e.g., Ruiz-Lapuente \& Canal 1998; Yungelson \& Livio 2000, and references therein).
The SN~Ia rate vs time in a given environment (e.g., cluster, or field)
will then be a convolution of the star-formation history in that environment 
with a ``delay'' or ``transfer function'', which is the SN~Ia 
rate vs time following a brief burst of star formation  [e.g., 
Sadat et al. 1998; Madau, Della Valle, \& Panagia 1998 (MDP); 
Dahlen \& Fransson 1999; Sullivan et al. 2000].

In the present paper, the cluster iron mass problem is revisited, 
and some simple relations connecting the iron mass and the SN rate 
to updated values of the various observables are derived. 
We then use the observed upper limits on the cluster
SN~Ia rate to set constraints on the SN~Ia progenitor models, 
on the formation of stellar populations in galaxy clusters,
and on the cluster enrichment scenario.
Throughout the paper we assume a flat cosmology with $\Omega_m=0.3$
and $\Omega_{\Lambda}=0.7$, and a Hubble parameter 
of $H_0=70$~ km s$^{-1}$ Mpc$^{-1}$. 

\section{The iron problem, and the SN rates needed to resolve it}

Detailed models of cluster metal enrichment have been calculated previously
(e.g., Buote 2002; Pipino et al. 2002; Finoguenov et al. 2003). 
However, the specific problem of the total iron mass in clusters
can be formulated rather simply as a function of observational parameters.
The ratio of observed iron mass to iron mass expected from
core-collapse SNe is
\begin{equation}
\frac{M_{\rm Fe-observed}}{M_{\rm Fe-SNII}}
=\frac{ M f_{\rm bar}(f_{gas}
 ~Z_{{\rm Fe-gas}}+f_* ~Z_{{\rm Fe}*})}   
      {M f_{\rm bar}~f_*~  f(>8 M_{\odot})~ f_{\rm Fe-SNII}}.
\end{equation}
Here $M$ is the mass of a cluster, $f_{\rm bar}$ is the baryon mass
fraction, $f_{gas}$ is the mass fraction of the baryons in the ICM, and
$Z_{\rm Fe-gas}$ is the mass fraction of that gas in iron. Similarly,
$f_*=1-f_{gas}$ is the mass fraction of the baryons in stars and $Z_{\rm Fe*}$ is
the iron abundance of the stars. 
In the denominator,
$f(>8 M_{\odot})$ is the ratio
of the initial stellar mass in stars of mass $>8 M_{\odot}$
(i.e., stars that underwent core-collapse), to the mass in stars
of lower mass, and  $f_{\rm Fe-SNII}$ is the
iron yield of core-collapse SNe, expressed as a fraction of the progenitor
masses.

Lin et al. (2003) have recently derived improved estimates of
stellar and gas mass fractions in clusters using infrared data from
the 2MASS survey. For rich clusters they find
$f_{gas}=0.9$ and $f_*=0.1$. Their stellar mass fraction is based
on the 2MASS K-band luminosity function (Kochanek et al. 2001)
combined with dynamical stellar mass-to-light ratio measurements
by Gerhard et al. (2001). Ettori (2003) has recently suggested that
a significant fraction, 6-38\%, of cluster baryons may be in a yet-undetected
warm gas. However, if this new component has a similar iron abundance
to that of the hot ICM, our arithmetic will not be affected. 
We adopt the ``canonical'' ICM iron abundance in rich clusters of
$Z_{\rm Fe-gas}\approx 0.3 ~Z_{\odot {\rm Fe}}$
(e.g., Mushotzky \& Loewenstein 1997;
Fukazawa et al. 1998; Finoguenov et al. 2000; White 2000).
This iron abundance relates to a photospheric
Solar iron  mass abundance of $Z_{\odot {\rm Fe}}=0.0026$ found  by 
Anders \& Grevesse (1989). [Using the updated photospheric Solar abundance
of $Z_{\odot {\rm Fe}}=0.00177$, given by 
Grevesse \& Sauval (1999), which also agrees with the meteoritic
Solar value of Anders \& Grevesse (1989), would imply simply raising the
ICM value accordingly]. 
Most of the stellar mass is in the elliptical galaxies, 
for which we adopt $Z_{\rm Fe*}\approx 1.2 ~Z_{\odot {\rm Fe}}$, 
the median found by J$\o$rgensen (1999) for early-type galaxies in Coma.

To estimate, $f(>8 M_{\odot})$, the ratio of exploding to non-exploding
initial stellar masses, an IMF, $dN/dm$, must be assumed:
\begin{equation}
f(>8 M_{\odot})=\frac
{\int_{8M_{\odot}}^{m_{up}}~dN/dm~ m~ dm}
{\int_{m_{low}}^{8M_{\odot}}~dN/dm~ m~ dm},
\end{equation}
where ${m_{low}}$ and ${m_{up}}$ are the lower and upper mass cutoffs
of the IMF. 
For a Salpeter (1955) IMF, $dN/dm\propto m^{-2.35}$.
Baldry and Glazebrook (2003) have recently modeled the local
UV-to-IR luminosity 
density of galaxies assuming a range of IMFs and SFHs, and found  
the data to be consistent with the Salpeter IMF.
A Salpeter slope with 
${m_{low}}=0.1 M_{\odot}$ and ${m_{up}}=100 M_{\odot}$, gives
$f(>8 M_{\odot})\approx 0.16$. However, other IMFs have been
proposed.
Figure 1 shows the dependence of
$f(>8 M_{\odot})$ on ${m_{low}}$ for single-power-law
IMFs (such as Salpeter's) of various indices $\alpha$, and 
for ``standard'' IMFs.
Varying ${m_{up}}$ has a weak effect on
$f(>8 M_{\odot})$, as long as the IMF is steep enough.
\begin{figure}
\centerline{\epsfxsize=120mm\epsfbox{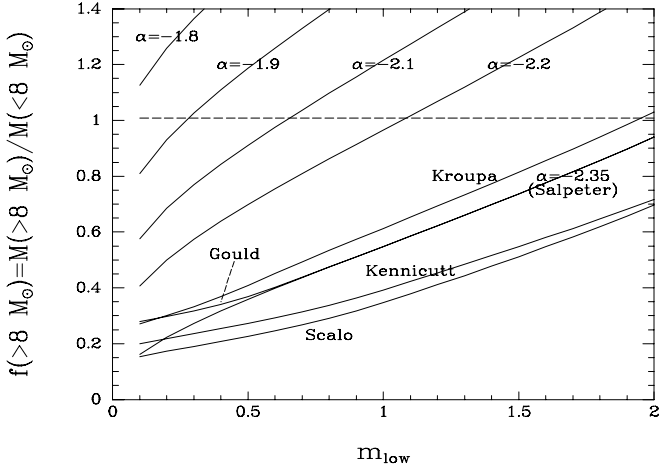}}
\caption{
The ratio
of the initial stellar mass in stars of mass $>8 M_{\odot}$
(i.e., stars that underwent core-collapse), to the initial mass in stars
of lower mass (i.e., the present-day cluster stellar mass), 
shown as a function of the lower mass cutoff
of the IMF, for various IMFs, as marked. The dotted line shows the
value of $f(>8 M_{\odot})$ needed to produce the observed 
iron mass in rich clusters with core-collapse SNe only, 
for the fiducial values of the other parameters
in Eq. \ref{feobssnii}. Single power-law IMFs (including Salpeter 1955)
are marked with their index. The ``standard'' non-top-heavy IMFs are
from Kennicutt (1983; $\alpha=-1.4,-2.5$ in the mass ranges $0.1-1M_{\odot}$,
$1-100M_{\odot}$, respectively); Gould et al. (1997; $\alpha
=-0.9,-2.21,-2.35$ for $0.1-0.6M_{\odot}$, $0.6-1M_{\odot}$,
$1-100M_{\odot}$); Scalo (1998; $\alpha
=-1.5,-2.7,-2.3$ for $0.1-1M_{\odot}$, $1-10M_{\odot}$,
$10-100M_{\odot}$); and Kroupa (2000; $\alpha
=-1.3,-2.3$ for $0.1-0.5M_{\odot}$, $0.5-100M_{\odot}$).  
}
\end{figure}  

Core collapse models generally agree on
an iron yield of about $0.1 M_{\odot}$ per SN (e.g., Thielemann, Nomoto,
\& Hashimoto 1996). In observed core-collapse SNe, estimates of 
the total yield of Ni$^{56}$ (which decays
to Fe$^{56}$ and drives the optical luminosity of SNe) have
been obtained by different methods -- the luminosity of the
radioactive tail, the luminosity of the plateau phase, and the 
H$\alpha$ luminosity in the nebular phase 
(Elmhamdi, Chugai, \& Danziger 2003). 
The different methods give consistent 
results for a given SN, but there is a large scatter among different SNe
(e.g., Zampieri et al. 2003) and a range of yields 
as large as 0.0016 to 0.26 $M_{\odot}$ (Hamuy 2003), 
with a mean of $0.05 M_{\odot}$ (Elmhamdi et al. 2003). 
Progenitor masses are more difficult
to estimate, but there are some indications that the iron yield and the
progenitor mass are correlated, i.e., the iron yield is some fraction of
the progenitor mass, of order 0.5-1\% (P. Mazzali, private communication). 
Since the ratio of the mean iron yield found by Elmhamdi et al. (2003; $0.05 M_{\odot}$)
to the minimum progenitor mass ($8 M_{\odot}$) is $0.063 \%$, and the
ratio is probably even smaller for those events with more massive
progenitors, we will err conservatively by assuming a large fractional iron
mass yield of $f_{\rm Fe~ SNII}\approx 0.01$.

Combining the estimates above, we can parametrize the iron problem as
\vspace{1cm}
\begin{eqnarray}
\frac{M_{\rm Fe-observed}}{M_{\rm Fe-SNII}}=
6.3
~\frac{\hspace{-.7cm}(f_{gas}~Z_{{\rm Fe-gas}}+f_* ~Z_{{\rm Fe}*})}{(0.9\times 0.3+0.1\times 1.2) 0.0026}
       \left(\frac{f_*}{0.1}\right)^{-1} \nonumber
\end{eqnarray}
\begin{equation}
\label{feobssnii}
\times~~    \left[\frac{f(>8 M_{\odot})}{0.16}\right]^{-1}
     \left(\frac{f_{\rm Fe-SNII}}{0.01}\right)^{-1}.
\end{equation}
For the fiducial values, the discrepancy is by a factor of about 6, 
as has been found by previous studies (e.g., Tozzi \& Norman 2001). 
 From Figure 1,
we see that none of the conventional IMFs can solve the iron problem,
which ranges from a factor 3.6 
[for a Gould, Bahcall, \& Flynn (1997) IMF] to a factor 6.6
[for a Scalo (1998) IMF] in the excess of the observed iron mass.
From the figure, we 
 can also see to what degree the IMF must be made ``top heavy'' in order
to solve the problem with core-collapse SNe only;
$m_{low}$ needs to be
greater than $2 M_{\odot}$, or $\alpha>-1.9$, or some other 
suitable combination of
$m_{low}>0.1M_{\odot}$ and $\alpha>-2.35$.

The discrepancy can be lowered somewhat 
by assuming a larger stellar mass fraction, $f_*$. 
However, only a completely unrealistic
value of $f_*\approx 1$, would lower the discrepancy to a non-crisis level.
Stated differently, the iron mass within the cluster galaxies is the
amount expected from core-collapse SNe and a normal IMF, while all the iron
in the ICM is an excess over this expectation. This, too, has been found
in detailed modeling (e.g. Brighenti \& Mathews 1998).

If the dominant contribution to the observed iron mass in clusters
is from SNe~Ia, it is straightforward to predict the rate of
these events, $R_{Ia}$, as a function of time $t$ or redshift $z$. 
As already noted, the SN~Ia rate vs time in a given environment
is the convolution of the star-formation history in that environment 
with a ``delay'' or ``transfer function'', $D(t)$, which is the SN~Ia 
rate vs time following a brief burst of star formation at $t=0$.
In the case of clusters, all evidence is that star formation occurred
considerably before $z\sim 1$. Let us assume, first, 
that star formation occurred in a brief burst, at cosmic time $t_f$,
corresponding to a redshift $z_f$. Then simply
\begin{equation}
R_{Ia}(t) \propto D(t-t_f).
\end{equation} 

Various authors have used different approaches to represent $D(t)$.
For example, Ruiz-lapuente \& Canal (1998) and Yungelson \& Livio (2000)
have attempted to derive physically motivated versions of $D(t)$ for the 
different progenitor scenarios, including binary evolution and
accretion physics. Such calculations are complex and, by necessity,
include a large number of assumptions and poorly known parameters.
Nevertheless, the resulting $D(t)$ functions have a number of generic
features: a delay until the progenitor population forms; a fast
rise to maximum; and a power-law or exponential decay. These general forms
suggest an alternative, more phenomenological, parametrization of $D(t)$,
as has been adopted by Sadat et al. (1998), MDP, and Dahlen \& Fransson (1999).

We follow the delay function parameterization given by MDP. 
It is assumed that the 
progenitors of SNe~Ia are white dwarfs, and therefore the overall time delay
includes the mass-dependent lifetime of the progenitor as a main-sequence
star, $\Delta t_{MS}$. Once the progenitor 
has become a white dwarf, it has a probability 
$\propto \exp(-{{\Delta t}\over \tau})$ 
to explode as a SN~Ia, where $\Delta t$ 
is the time since the star left the main sequence.
Following MDP, then 
\begin{equation}
\label{mdpdelay}
D(t) \propto \int_{m_{\rm min}(t)} 
^{m_{\rm max}} \exp(-{t-\Delta t_{MS}\over \tau})\frac{dN}{dm}dm .
\end{equation}
For consistency, a Salpeter (1955) IMF is assumed. 
The minimum and maximum initial masses that 
will lead to the formation of a WD that explodes as a SN~Ia are
\begin{eqnarray}
m_{\rm min}={\rm max}[3 M_{\odot}, 
({{t-t'}\over{10~{\rm Gyr}}})^{-0.4} M_{\odot}],~~~~ 
m_{\rm max}=8 M_{\odot}, \nonumber
\end{eqnarray}
and
\begin{equation}
{{\Delta t_{MS}} \over {10~{\rm Gyr}}}=({{m} \over {M_{\odot}}})^{-2.5}.
\end{equation}
Figure 2 shows several examples of the MDP delay function.
After an instantaneous starburst\footnotemark[1] at $t=0$, the function is zero
\footnotetext[1]{See MDP, Eq. 4, for the treatment of non-instantaneous star formation.} 
for 55~Myr (until the first white dwarfs form), 
then rises approximately as $t^{0.5}$, and reaches
a peak at $t=0.64$~Gyr. It then declines exponentially with a timescale 
$\tau$.

\begin{figure}
\centerline{\epsfxsize=140mm\epsfbox{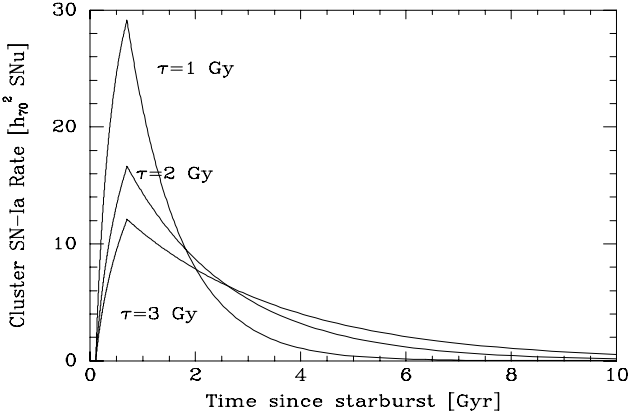}}
\caption{Examples of model SN~Ia delay functions, following the
prescription of MDP. All curves are normalized to have the same area,
i.e., to give a time-integrated total number of SNe~Ia that will 
produce the observed iron-mass-to-light ratio of rich clusters.}
\end{figure}  

In a given cluster, the integral on $R(t)$ over time gives the total 
number of SNe~Ia that have exploded in that cluster.
If most of the cluster iron is from SNe~Ia, 
this total number is just the observed iron mass
(minus the small fraction
expected from core-collapse SNe arising from a normal IMF, Eq.
\ref{feobssnii}), divided
by the iron mass yield per SN~Ia.
The normalization of $R(t)$ is therefore set by
\begin{equation}
\label{intria1}
\int R_{Ia}(t)~dt=\frac{M_{\rm Fe-observed}}
               { m_{\rm Fe-Ia}}
               (1-\frac{M_{\rm Fe-SNII}}{M_{\rm Fe-observed}}).
\end{equation} 

The mean iron yield of a single SN~Ia, $m_{\rm Fe-Ia}$,
is generally agreed
to be about $0.7\pm 0.1 M_{\odot}$. 
This emerges from modeling of the bolometric
light curves of SNe~Ia (e.g., Contardo, Leibundgut, \& Vacca 2000,
and references therein), as well as from SN~Ia model calculations 
(e.g., Thielemann, Nomoto, \& Yokoi 1986). 

If star formation is not in an instantaneous burst, but rather begins
at $t_f$ and lasts some period of time, the occurrence of some SNe~Ia
will be delayed. The SN~Ia rate at later 
times, given the same total iron mass,
will be necessarily greater than under the brief burst assumption.
The assumption of a single, brief, star-formation burst at $t_f$ will therefore
lead to a lower limit for the predicted SN~Ia rate at later times.

Rewriting Eq. \ref{intria1} with the
fiducial values above and in Eq. \ref{feobssnii}, 
and normalizing the SN rate by the present-day $B$-band stellar luminosity
of the cluster, we obtain
\begin{eqnarray}
\frac{1}{L_B}\int R_{Ia}(t)~dt = \nonumber
\end{eqnarray}
\begin{eqnarray}
0.042\frac{\rm SN}{L_{B\odot}}
\left(\frac{M/L_{B}}{200}\right) \left(\frac{f_{\rm bar}}{0.17}\right) 
~\frac{\hspace{-.7cm}(f_{gas}~Z_{{\rm Fe-gas}}
+f_* ~Z_{{\rm Fe}*})}{(0.9\times 0.3+0.1\times 1.2) 0.0026} \nonumber
\end{eqnarray}
\begin{equation}
\label{intria2}
\times ~~ \left(\frac{m_{\rm Fe-Ia}}{0.7 M_{\odot}}\right)^{-1} ~~ \times ~~
\frac{1-(\frac{M_{\rm Fe-observed}}{M_{\rm Fe-SNII}})^{-1}}{5/6}.
\end{equation}
Here we have used a value of 
$M/L_B=(200 ~\pm 50) \frac{M_{\odot}}{L_{B\odot}}$ 
for the typical
total mass-to-light ratio measured in rich clusters, assuming  
$H_0=70$~km s$^{-1}$ Mpc$^{-1}$. The central value and the error is 
obtained by taking the union of several recent determinations
and their quoted uncertainties --
Carlberg et al. (1996), Girardi \& Giuricin (2000),
Bahcall \& Comerford (2002), and Girardi et al. (2002). 
Galaxy kinematics, modeling of the X-ray emission,
and strong and weak gravitational
lensing give generally consistent results for this parameter.
The fraction of the cluster mass that is in baryons, $f_{\rm bar}$, 
is 10-30\% (Ettori \& Fabian 1999; Mohr, Mathiesen \& Evrard 1999;
Allen et al. 2002; Arnaud et al. 2002), 
and is thought to be representative
of the universal baryonic mass fraction. We therefore adopt the
cosmic value, 0.17, as recently measured with WMAP (Spergel et al. 2003).
 
The time-integrated number of SNe~Ia per unit stellar luminosity
obtained from Eq. \ref{intria2} for the fiducial values, can also be expressed as
\begin{equation}
0.042 \frac{\rm SN}{L_{B\odot}}=42~{\rm SNu~Gyr}.  
\end{equation}
In other words, to produce the 
iron mass seen in clusters with SNe~Ia, there must have been in the past
one SN~Ia for every $23L_{B\odot}$ of present-day stellar luminosity.
Equivalently, the mean SN~Ia rate over a $\sim 10$~Gyr cluster age must have
been 4.3~SNu. Since present day rates (e.g., in Virgo cluster ellipticals;
Cappellaro et al. 1999) are much
lower than this, the rate must have been much higher in the past. 

Thus, the assumption of a brief star-formation burst at some time $t_f$ 
in the past, followed by a SN~Ia rate $D(t-t_f)$ with characteristic
time $\tau$, determines the form of $R_{Ia}(t)$. 
The observed iron mass determines the normalization of $R_{Ia}(t)$.
Extended or multiple starbursts after $t_f$ can only lead to a higher
$R_{Ia}(t)$ (except at times very soon after $t_f$).
The only free parameters are therefore $t_f$ and $\tau$,
which can be constrained by comparison to direct measurements of $R_{Ia}(t)$.
Time and redshift are related, for our chosen cosmology, by
\begin{equation}
\Delta t=H_0^{-1} \int_{z_1}^{z_2}(1+z)^{-1}[\Omega_m (1+z)^3+\Omega_{\Lambda}]^{-0.5} dz.
\end{equation}

\section{Comparison to Observations and Conclusions}

Figure 3 shows the expected $R_{Ia}$ vs. redshift based on Eqns. \ref{mdpdelay}
and \ref{intria2}, for 
several choices of $\tau$ and cluster stellar formation redshift 
$z_f$, and compares these predictions to the observations.
Existing measurements of cluster SN~Ia rates are by Reiss (2000)
for $0.04<z<0.08$, and by Gal-Yam et al. (2002) 
for $z=0.25^{+0.12}_{-0.07}$, and $z=0.90^{+0.37}_{-0.07}$. 
The error bars show the 95\% confidence intervals.
Figure 4 is the same, but zooms in on the region $z<1.4$, where the data exist.

\begin{figure}
\centerline{\epsfxsize=140mm\epsfbox{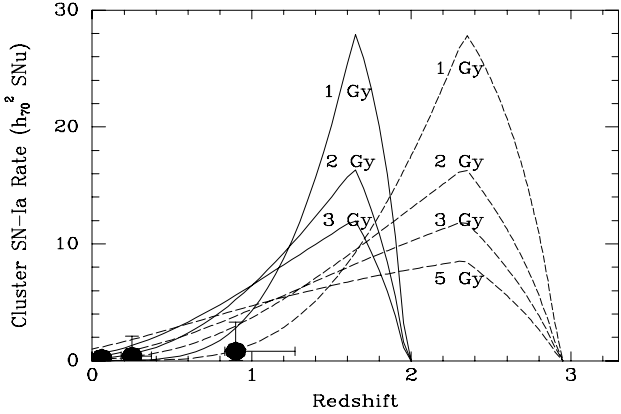}}
\caption{Predicted SN~Ia rates vs redshift, if most of the iron mass
in  clusters is produced
by type-Ia SNe following a brief burst of star formation at redshift  $z_f=2$
(solid curves) and $z_f=3$ (dashed curves). The different curves are
for SN~Ia transfer functions with mean delay times,
$\tau$, as marked. Cluster SN~Ia rate measurements are
by Reiss (2000) and Gal-Yam et al. (2002). The latter are shown
with 95\%-confidence vertical error bars.  The horizontal error bars
give the visibility-time-weighted redshift ranges of the cluster samples.}
\end{figure}  

\begin{figure}
\centerline{\epsfxsize=140mm\epsfbox{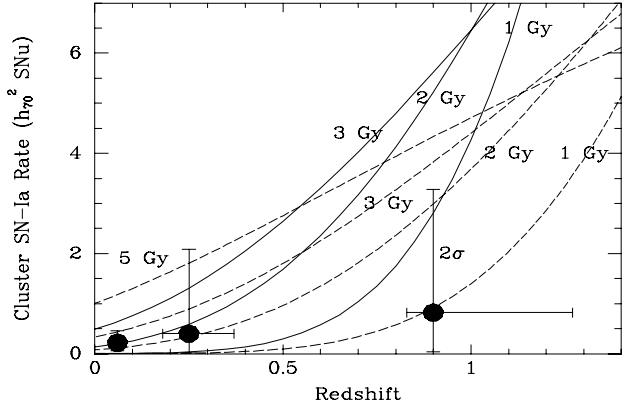}}
\caption{Same as Fig. 3, but zooming in on $z<1.4$.
The $z_f=2$ models (solid curves) with $\tau\ge 2$~Gyr are clearly 
ruled out by the $z\sim 1$ SN-rate measurement, even after accounting
for a 30\% uncertainty in the nomalization of the models.
The $z_f=3$ model (dashed curves) with $\tau=5$~Gyr predicts unacceptably
high rates at low $z$.} 
\end{figure}

Figure 4 shows that the $z\sim 1$ SN~Ia rate measurement is inconsistent
with several of the plotted models. However, in the comparison, we must keep in
mind that there are uncertainties in the parameters entering
Eq. \ref{intria2}, which sets the normalization of the curves.
The main uncertainty, $\pm 25\%$, is in the mean $M/L_{B}$ of clusters.
Accounting also for the (smaller) uncertainties in the other parameters,
we consider the measured
95\% upper limit on $R_{Ia}$ at $z=0.9$ to be in conflict with the model
with $z_f=2$, $\tau=2$~Gyr, and the model 
with $z_f=2$, $\tau=3$~Gyr.
The upper limit on $R_{Ia}(t)$ is
at $\sim 60\%$ of the predicted values for these models.
The model with $z_f=3$, $\tau=5$~Gyr
is marginally consistent with the $z=0.9$ data point, given the
uncertainty in the prediction. However, this model gives an unacceptably
high rate at low redshift, and therefore can also be rejected.
The low observed SN~Ia rate at $z\sim0-1$
means that, if cluster iron was produced by SNe~Ia, those SNe
must have occurred at earlier times, times that have yet to be probed 
by observations. To push the SNe~Ia to such early times, one must invoke
early star formation {\it and} a short SN~Ia delay time.

A recent attempt by van Dokkum \& Franx (2001)
to deduce the epoch of star formation in
cluster ellipticals by spectral synthesis modeling of the
stellar populations, and accounting for biases in the
selection of elliptical galaxies,
finds stellar formation redshifts of $z_f=2^{+0.3}_{-0.2}$.
 If we adopt a particular formation
redshift, say $z_f=2$, the resulting upper limit of $\tau<2$~Gyr places
clear constraints on SN~Ia progenitor models. Two of the models
by Yungelson \& Livio (2000) predict longer delays and are therefore
ruled out. In the double degenerate model, two
WDs merge, and the resulting object, having a mass larger
than the Chandrasekhar limit, is subject to a runaway thermonuclear
explosion triggered at its core. Yungelson \& Livio find that the delay 
function of such systems has an exponential 
cutoff at $\sim11$ Gyr, at odds with the above upper limit on $\tau$. 
A similar discrepancy exists for a model where a WD accrets He-rich 
material from a He-star companion, leading to helium ignition on the
surface of the WD and an edge-lit detonation of the star. This model
has a delay function that cuts off at $\sim 5$~Gyr.
    
One must recap here that all these conclusions hold under the assumption
that 
the stars in galaxy clusters were formed with a standard IMF, and therefore
most of the iron is from SNe~Ia. If the stars were formed with a 
sufficiently top-heavy
IMF to produce the observed iron, few SNe~Ia are expected
at any redshift, and the SN~Ia rate measurements place no constraints
on cluster star-formation epoch or on SN~Ia time delay.  

Since our conclusions are based on the comparison of a low observed
cluster SN~Ia rate at $z\sim 1$ to the high rate predicted by some
models, it is sensible to re-examine the reliability of the observation.
One possibility to consider is that the rate measured by Gal-Yam et al. (2002), 
based on deep {\it HST} cluster images,
was low because the SN detection efficiency was overestimated. 
Some faint cluster SNe that were missed
would then be incorrectly accounted for in the rate calculations.
This is highly unlikely, given that the actually detected 
cluster SNe were relatively bright, with $I<24$~mag. Much fainter 
SNe, down to 28~mag, {\it were} found by the survey, demonstrating its
high sensitivity, but these SNe were background and foreground events, rather
than cluster events. A second possibility is that the rate measured
was skewed low by considering only the visibility
time of normal SNe~Ia. Ignoring a significant population of subluminous
SNe~Ia, which have short visibility times, will lower the derived rate.
However, studies of local subluminous SNe~Ia show that they have an 
extremely low iron yield, $\sim 0.007 M_{\odot}$ (e.g., Contardo et al.
2000). Thus, unless
they are very common at $z\sim 1$, such SNe are irrelevant for
iron production in clusters.
 
An interesting conclusion arises if we examine our results jointly
with the results we report in a companion paper (Gal-Yam \& Maoz 2003). 
There, we find that some versions of cosmic SFH,
combined with particular SN~Ia delay times, are incompatible with the
observed redshift distribution of SNe~Ia found by Perlmutter et al. (1999).
For example, if the cosmic SFR rises between $z=0$ and $z\sim 1$ as sharply
as implied by Lilly et al. (1996) and by Hippelein et al. (2003), 
and at $z>1$ as sharply as found by Lanzetta et al. (2002), then the
observed SN~Ia
redshift distribution sets a 95\%-confidence {\it lower} limit on the
SN~Ia delay time, of
$\tau>3$~Gyr. Thus, if one had independent reasons to adopt
this SFH scenario, {\it and} a cluster star-formation redshift $z_f<2$
(the latter 
implying an {\it upper} limit of $\tau<2$~Gyr at 95\% confidence
from the measured SN~Ia rate), then {\it all} values of
$\tau$ would be excluded. One would then be
forced to the conclusion that SNe~Ia cannot be the source of iron in
clusters, leaving only the top-heavy IMF option.

To summarize, we have investigated the source of the total iron mass
in rich galaxy clusters. Using updated values for the various observational
parameters, we have rederived and quantified the excess of iron over
the expectation from core-collapse SNe, provided the stars in cluster galaxies
formed with a standard IMF. Assuming the source of the iron is from SNe~Ia,
we then showed that the SN~Ia rate vs. redshift can be predicted quite
robustly, given two parameters -- the star formation redshift in clusters,
and the mean SN~Ia delay time. We set constraints on these two parameters
by comparing the predicted rates to the measurements of the SN~Ia rate
at $z\sim 1$ by Gal-Yam et al. (2002). 
The low observed rates at $z\sim 1$ force
the iron-producing SNe~Ia to have occurred at higher redshifts. This 
implies an early epoch of star formation {\it and} a short delay time.
Specifically, we showed that models with a mean SN~Ia delay time of 
$\tau<2$~Gyr ($<5$~Gyr) and $z_f<2$ ($z_f<3$) are ruled out at high confidence.
Thus, if other avenues of inquiry show that SNe~Ia explode via a mechanism that
leads to a long delay time that is ruled out by our study, it will
mean that core-collapse SNe from a top-heavy IMF must have formed the
bulk of the observed mass of iron in clusters. The same conclusion will apply
if other studies confirm a cosmic SFH that rises sharply with redshift.
As shown in Gal-Yam \& Maoz (2003), short ($<4$~Gyr) delay times are
incompatible with such a SFH and the
redshift distribution of field SNe~Ia. 

Finally, we note that 
our results have relied on the upper limits
on the $z \sim 1$ cluster SN~Ia rate, 
calculated by Gal-Yam et al. (2002) based on just several SNe.
Improved measurements of cluster SN rates at low, intermediate,
and high redshifts can tighten
the constraints significantly, and can potentially reveal the iron source
directly.     
  
\section*{Acknowledgments}

We thank M. Hamuy,
A. Filippenko, P. Mazzali, and B. Schmidt for valuable comments and
discussions. This work was supported by the Israel Science Foundation --- the
Jack Adler Foundation for Space Research, Grant 63/01-1.

\clearpage


\end{document}